\documentclass[reprint,preprintnumbers,amsmath,amssymb,aps,nofootinbib]{revtex4-1}
\usepackage[pdftex]{graphicx,hyperref}
\usepackage{dcolumn,bm,cases,natbib}
\usepackage[svgnames]{xcolor}
\definecolor{phthaloblue}{rgb}{0.0, 0.06, 0.54}
\hypersetup{
    colorlinks=true,
    linkcolor=phthaloblue,
    citecolor=blue,
    filecolor=blue,
    urlcolor=phthaloblue,
    }
\makeatletter \def\@eqnnum{{\normalsize \normalcolor (\theequation)}}  \makeatother %

\def\({\left(}
\def\){\right)}
\def\[{\left[}
\def\]{\right]}

\def\nn{\nonumber \\}

\def\lmk{\left(}
\def\rmk{\right)}
\def\lkk{\left[}
\def\rkk{\right]}
\def\dd{{ d}}
\def\la{\left<}
\def\ra{\right>}
\def\del{\partial}
\newcommand{\eq}[1]{Eq.~(\ref{#1})}
\newcommand{\beq}{\begin{eqnarray}} 
\newcommand{\eeq}{\end{eqnarray}}
\newcommand{\bel}[1] {\begin{equation}\label{#1}}
\newcommand{\beal}[1] {\begin{eqnarray}\label{#1}}
\newcommand{\be}{\begin{equation}}
\newcommand{\ee}{\end{equation}}
\newcommand{\bea}{\begin{eqnarray}} 
\newcommand{\eea}{\end{eqnarray}}
\newcommand{\abs}[1]{\left\vert#1\right\vert}

\newcommand{\Ps}{\mathcal{P}_{\zeta}}
\newcommand{\ns}{n_s}

\begin{document}

\title{Primordial Black Holes from Polynomial Potentials in Single Field Inflation}

\author{Mark P. Hertzberg}
\email{mark.hertzberg@tufts.edu}
\affiliation{Institute of Cosmology, Department of Physics and Astronomy, 
Tufts University, Medford, MA  02155, USA}

\author{Masaki Yamada}
\email{masaki.yamada@tufts.edu}
\affiliation{Institute of Cosmology, Department of Physics and Astronomy, 
Tufts University, Medford, MA  02155, USA}

\begin{abstract}
Within canonical single field inflation models, we provide a method to reverse engineer and reconstruct the inflaton potential from a given power spectrum. This is not only a useful tool to find a potential from observational constraints, but also gives insight into how to generate a large amplitude spike in density perturbations, especially those that may lead to primordial black holes (PBHs). In accord with other works, we find that the usual slow-roll conditions need to be violated in order to generate a significant spike in the spectrum. We find that a way to achieve a very large amplitude spike in single field models is for the classical roll of the inflaton to over-shoot a local minimum during inflation. We provide an example of a quintic polynomial potential that implements this idea and leads to the observed spectral index, observed amplitude of fluctuations on large scales, significant PBH formation on small scales, and is compatible with other observational constraints. We quantify how much fine-tuning is required to achieve this in a family of random polynomial potentials, which may be useful to estimate the probability of PBH formation in the string landscape.
\end{abstract}

\date{\today}
\maketitle
\flushbottom
\allowdisplaybreaks[1]

\section{Introduction}
\label{sec:intro}

Cosmological observations of the CMB and large scale structure are compatible with the theory of early universe inflation \cite{Guth:1980zm}. Inflation naturally produces an approximately scale invariant spectrum of small density perturbations on large scales, which also extends to small scales in the simplest models. However, recently there has been increasing interest in the possibility of breaking scale invariance on small scales with a spike in the power spectrum. A spike in the primordial power spectrum is interesting as it may lead to structures such as primordial black holes (PBHs)~\cite{Hawking:1971ei, Carr:1974nx, Carr:1975qj, Ivanov:1994pa}. 

The idea of filling the universe with PBHs is of interest for several reasons. One reason is that PBHs could be a significant component of the dark matter in the universe. Although there are several constraints on the energy density of BHs and most recent analyses suggest PBHs are unlikely to be all the dark matter for generic masses, there may be a small window for PBHs to be a significant component (or all) of the dark matter near the mass scale  $\sim 10^{20} \, {\rm g}$~\cite{Inomata:2017okj} (much lighter or much heavier PBHs are of interest too). A second reason is that even if PBHs are only a small fraction of the energy budget of the universe, they could still lead to interesting astrophysical phenomena, such as acting as a seed for the production of super massive black holes at the center of galaxies~\cite{Rubin:2001yw, Bean:2002kx, Duechting:2004dk}, or producing binary mergers and gravitational waves~\cite{Bird:2016dcv, Clesse:2016vqa, Sasaki:2016jop, Carr:2016drx}. 

Such heavy PBHs can form if there are large amplitude primordial scalar perturbations on the relevant scale~\cite{Ivanov:1994pa}. Depending on the statistical distribution, the tail of the distribution will lead to highly over-dense regions at random places in the universe. For sufficiently large over-densities, these will collapse as they re-enter the horizon in the radiation-dominated era, leading to the formation of a black hole~\cite{Hawking:1971ei, Carr:1974nx, Carr:1975qj}. 

The recent detection of binary black hole (BH) mergers by the Advanced LIGO and Advanced Virgo detectors involved BHs in the mass range of $8 - 35 M_\odot$~\cite{Abbott:2016blz, Abbott:2016nmj, Abbott:2017vtc, Abbott:2017oio}. Some of these GW events could be a merger of two PBHs~\cite{Bird:2016dcv, Clesse:2016vqa, Sasaki:2016jop, Carr:2016drx}. The merger rate can be in agreement with the one suggested by observations if the fraction of PBHs in dark matter is of order $10^{-3}$, and it was claimed that this is consistent with existing observational upper bounds (see, e.g., Refs.~\cite{Carr:2017jsz, Inomata:2017uaw, Kocsis:2017yty}). 

Such an exotic interpretation requires a correspondingly large amplitude of power spectrum at small scales. However, there are many constraints on the amplitude of the primordial power spectrum (see, e.g., Ref.~\cite{Bringmann:2011ut}). The observations of CMB temperature anisotropies provide a precise measurement of the power spectrum at large scales~\cite{Nicholson:2009pi, Nicholson:2009zj}. The large scale structure and the Lyman-alpha forest observations give us information on slightly smaller scales~\cite{Bird:2010mp}. The $\mu$-distortion~\cite{Chluba:2012we} and the secondary gravitational effect~\cite{Inomata:2016rbd} give other severe constraints on the amplitude of primordial perturbation for smaller scales. If the over-density is beyond a certain threshold, a large density region called an ultra-compact mini-halo (UCMH) will form after matter-radiation equality. If the main component of dark matter is a weakly interacting massive particle (WIMP), the annihilation in an UCMH can produce highly luminous gamma-ray or neutrino sources.\footnote{UCMHs lead to a time delay in pulsar timing by the Shapiro effect and the PTA experiments put an upper bound on the amplitude of power spectrum~\cite{Clark:2015sha, Emami:2017fiy} (see also Refs.~\cite{Siegel:2007fz, Schutz:2016khr}). However, in Ref.~\cite{Clark:2015sha} they claimed that they used an optimistic value for the sensitivity of pulser timing experiment. The constraint is absent when we use a realistic value, so that we do not consider it in this paper.} The constraints on these fluxes also give severe upper bounds on the power spectrum, though they depend sensitively on the details of the dark matter models~\cite{Lacki:2010zf, Bringmann:2011ut, Nakama:2017}. A large scalar perturbation can generate gravitational waves via second-order effects, which would affect pulsar timing~\cite{Saito:2008jc, Saito:2009jt}. This gives a model-independent constraint in the range of $k \sim 10^{6 - 7} \, {\rm Mpc}^{-1}$. 

Hence any inflation model that predicts large amplitude fluctuations over some range of scales must be compatible with these constraints. There are many inflation models that can generate such a large amplitude spike in the power spectrum~\cite{Garcia-Bellido:2017mdw, Ezquiaga:2017fvi, Kannike:2017bxn, Germani:2017bcs, Motohashi:2017kbs, Ballesteros:2017fsr} (see Refs.~\cite{GarciaBellido:1996qt, Kawasaki:1997ju, Yokoyama:1998pt, Kawaguchi:2007fz, Kohri:2007qn, Frampton:2010sw, Drees:2011yz} for earlier works). The most well known is in hybrid inflation, which involves multiple fields undergoing a tachyonic phase transition; though it is somewhat difficult to achieve the correct spectral index on CMB scales in these models (e.g., see Ref.~\cite{Halpern:2014mca}). Achieving a spike in the power spectrum on small scales within the framework of a canonical two-derivative single field inflationary model is somewhat difficult to achieve, but will be the focus of our work in this paper. In this case, a very small slope in the potential is generally needed to generate a large fluctuation. Some existing approaches focus on an inflection point inflation combined with an inflation model that predicts a consistent spectral index. In other words, they focus on a known inflation model and modify it so as to reach an inflection point before the end of inflation and give a large amplitude at a small scale. 

In this paper we begin by developing a method to reconstruct inflaton potentials from an arbitrary power spectrum within the framework of single field two-derivative actions; this acts to ``reverse engineer" the model from the data. This is not only a useful tool to find a potential from observational constraints, but also gives insight into how to generate large amplitude perturbations that break scale invariance. In particular, we find that the slow-roll condition has to be violated~\cite{Garcia-Bellido:2017mdw, Ezquiaga:2017fvi, Kannike:2017bxn, Germani:2017bcs, Motohashi:2017kbs, Ballesteros:2017fsr} and we can achieve this by having the inflaton overshoot a local minimum during inflation~\cite{Germani:2017bcs, Ballesteros:2017fsr}. Equipped with these facts, we then provide a polynomial potential that leads to the observed spectral index and significant PBH formation. We estimate the level of fine-tuning such random potentials need in order to generate PBHs in that are both appreciable and compatible with current constraints. These types of random potentials may be representative of the type of phenomena that can emerge in the string landscape (see, e.g., Ref.~\cite{Masoumi:2016eag}).

This paper is organized as follows:
In Section \ref{sec:PBH} we review the calculation of PBH formation from a primordial perturbation and the constraints on the power spectrum. 
In Section \ref{sec:formula} we explain the method to reconstruct inflaton potentials from a given power spectrum. 
In Section \ref{sec:applications} we then apply this method to some examples, including the one that predicts PBH dark matter with $M \sim 10^{20} \, {\rm g}$. We show that the slow-roll conditions need to be violated and the inflaton may overshoot a local minimum during inflation in order to generate a large amplitude of power spectrum. 
In Section \ref{sec:polynomial} we provide an explicit quintic polynomial potential that predicts a consistent spectral index with the observed value and a spike in the power spectrum as large as $10^{-2}$ at small scales, and we quantify the level of fine-tuning required to achieve this. 
Finally, we conclude and discuss our results in Section \ref{sec:conclusion}.

\section{PBH formation from primordial fluctuations}
\label{sec:PBH}

In this section we briefly review the formation and constraints on PBHs from primordial fluctuations. For brevity, we will not go into the full details, but follow a simple conventional analysis 
to illustrate the situation.

\subsection{PBH Abundance}

Suppose that a large amplitude of primordial fluctuation is generated at a small scale during inflation. After inflation ends, the comoving horizon increases and the mode re-enters the horizon at a later time. If on some scales the amplitude of primordial fluctuations is significantly large, there will be some number of over-dense regions where the density of matter is so large that it collapses to form a PBH upon horizon re-entry. For the PBH masses of interest this collapse will occur in the radiation era.

The mass of the resulting PBHs is comparable to the horizon mass at the time of re-entry: 
\beq
M(k)  &=& \left. \gamma\, \rho\, \frac{4 \pi H^{-3}}{3} \right\vert_{k = aH}, \nn
	&\simeq& 	10^{20}\mathrm{g} \left(
    \frac{\gamma}{0.2}
    \right)\!
    \left(
    \frac{g_\ast}{106.75}
    \right)^{\!- \frac{1}{6}}\!
    \left(
    \frac{k}{7 \times 10^{12} \,\textrm{Mpc}^{-1}}
    \right)^{\!-2} \!\!\!\!\!\!,\,\,\,\,\,\,\,\,\,
\label{eq:pbhmass}
\eeq
where $\gamma$ ($\sim 0.2$) is a proportionality constant~\cite{Carr:1975qj} and $g_*$ is the effective number of relativistic degrees of freedom. The ratio of the relevant wavenumber $k$ and the pivot scale $k_*$ ($= 0.05 \ {\rm Mpc}^{-1}$) is given by 
\beq
 \ln \lmk \frac{k}{k_*} \rmk \simeq 33 - \frac{1}{2} \ln \lmk \frac{M (k)}{10^{20} \, {\rm g}} \rmk, 
\eeq
where we assumed $\gamma = 0.2$ and $g_* = 106.75$. 

Assuming that the primordial fluctuations are Gaussian, we can provide a rough estimate for the fraction of PBHs formed with mass $M$ compared to the total radiation energy density as follows
\beq
	\beta (M) 
	&\equiv& \frac{\rho_{\rm PBH}(M)}{\rho_{\rm tot}} 
	\approx \int_{\delta_c}^\infty 
	\frac{\dd \delta}{\sqrt{2 \pi \sigma^2 (M)}} \, e^{- \frac{\delta^2}{2 \sigma^2(M)}},
	\\
	&\simeq&
	\sqrt{\frac{2}{ \pi}} \frac{ \sigma (M)}{\delta_c} \, e^{- \frac{\delta_c^2}{2 \sigma^2(M)}},\label{betaapp} 
\eeq
where $\delta_c$ ($\sim 0.3$) is the threshold of density perturbation above which a PBH forms at the time of re-entry~\cite{Carr:1975qj}. In the second line we have used the fact that it is only fluctuations in the upper tail of the distribution that form black holes in order to approximate the error function. Here the variance $\sigma^2 (M)$ is the coarse-grained variance of density perturbations smoothed on a scale $R = 1/k$. During the radiation dominated era, it is given by~\cite{Young:2014ana}
\beq
 \sigma^2 (M(1/R)) = \frac{16}{81} \int \dd \ln k' \lmk k' R \rmk^4 {\cal P}_{\zeta} (k') \, W(k' R)^2, \,\,\,
\label{sigmasquared}\eeq
where ${\cal P}_\zeta$ is the power spectrum of the primordial comoving curvature perturbations. Here $W(x)$ is a smoothing window function. Its form is often taken to be a Gaussian $W (x) = \exp \lmk - x^2 / 2 \rmk$ in the literature. Note that the factor of ($4\,k'^2R^2/9)^2$ essentially comes from the (relativistic version of) Poisson equation to convert curvature perturbations to density perturbations.

In order to determine the total energy density in PBHs, we integrate over all masses $M$ as follows
\beq
	\frac{\Omega_{\text{PBH,tot}}}{\Omega_c} = \int \dd \ln M\, \frac{\Omega_\text{PBH}(M)}{\Omega_c }, 
\eeq
where $\Omega_c h^2$ ($\simeq 0.12$) is the total dark matter abundance. The conversion from the energy density fraction in the radiation era to the fraction of dark matter today can be obtained by simple red-shifting as 
\beq
	&&\frac{\Omega_\text{PBH}(M)}{\Omega_c }
	= \lmk \frac{T_M}{T_\text{eq}} \frac{\Omega_mh^2}{\Omega_ch^2} \rmk \gamma\, \beta (M) ,  \nn
	&&\simeq
	\lmk \frac{\beta (M)}{8 \times 10^{-15}} \rmk 
	\lmk\frac{\gamma}{0.2} \rmk^\frac{3}{2}
	\lmk \frac{106.75}{g_{\ast} } \rmk^\frac{1}{4}
	\lmk \frac{M}{10^{20} \ {\rm g}} \rmk^{-\frac{1}{2}}. \,\,\,
\eeq
Here, $T_M$ is the temperature at the time of re-entry and $T_{\rm eq}$ is the temperature at matter-radiation equality. 

Hence an appreciable abundance of PBHs on some mass scale $M$, say within a few orders of magnitude of $10^{20}$\,g, requires $\beta(M)$ to be within a couple of orders of magnitude of $10^{-15}$, or so. For much heavier PBHs, such as $M\sim 30\,M_\odot$, then $\beta(M)$ should be $\lesssim 10^{-7}$ to avoid over-closure and may in fact need to be a couple orders of magnitude smaller to be compatible with other observational constraints. In any case, since Eq.~(\ref{betaapp}) is exponentially sensitive to $\sigma^2(M)$, then in order to have appreciable PBH production $\sigma^2(M)$ should be on the order $\sim 10^{-2}$, or so. This in turn arises from a power spectrum ${\cal P}_\zeta$ in Eq.~(\ref{sigmasquared}) that is also on the order of $10^{-2}$, or so, on the relevant scale.

\subsection{Constraints on Power Spectrum}

There are various constraints on the abundance of BHs. We will neglect the mass growth of PBHs via the accretion of matter onto PBHs after formation because we expect most of the PBHs remain in dark matter halos, 
whose density is too low to accrete onto these relatively small black holes. Hence we can directly relate the constraints on the abundance of PBHs $\Omega_{\text{PBH}}(M)$ to constraints on the primordial matter power spectrum ${\cal P}_\zeta$.

\begin{figure}[t] 
\centering
\includegraphics[width=8cm,height=7cm]{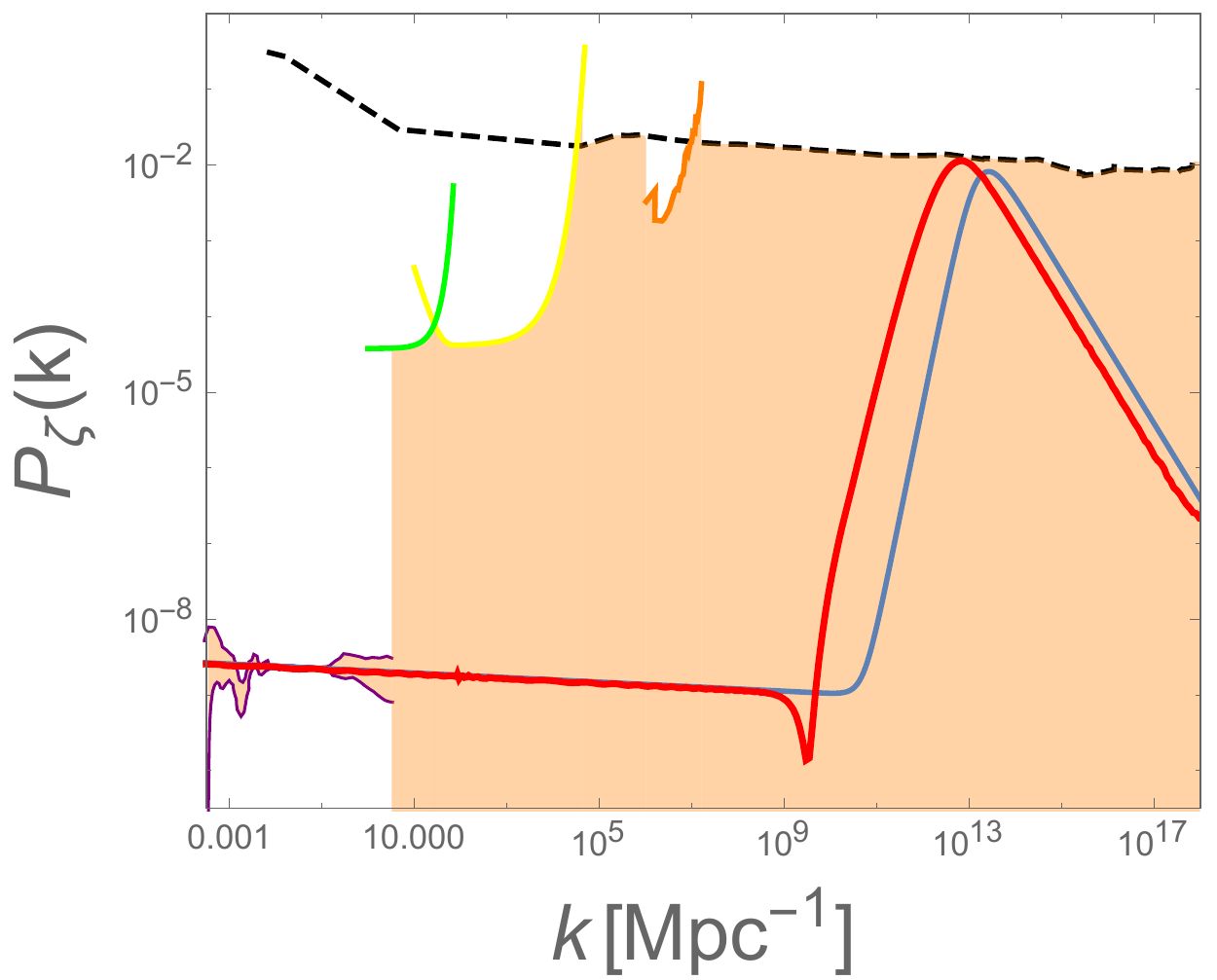} 
\caption{Example of power spectrum (blue line) that is put in by hand so that it produces a significant number of PBHs and is consistent with present constraints (shaded region is allowed): CMB and LLS survey (purple)~\cite{Nicholson:2009pi, Nicholson:2009zj, Bird:2010mp, Bringmann:2011ut}, $\mu$ (yellow) and $y$-distortions (green)~\cite{Chluba:2012we}, pulsar timing (orange)~\cite{Lentati:2015qwp, Shannon:2015ect, Arzoumanian:2015liz}, and PBH constraints (black dashed). In Section \ref{sec:app2}, we approximately reconstruct a potential from this blue input power spectrum. To check the consistency of our method, we take the approximate reconstructed potential and then numerically compute the power spectrum (red line) from it exactly.}
\label{fig:constraint}
\end{figure}

As a convenient way of parameterizing a peak (spike) in the power spectrum, with a peak on a scale $\sim k_p$, we assume the following form of the power spectrum:
\beq
 {\cal P}_\zeta (k) = \text{Max}\left\{\frac{4A}{\lkk \lmk k_p / k \rmk^a + \lmk k / k_p \rmk^b \rkk^2}\,,\,\mathcal{P}_\zeta^{\text{SI}}(k)\right\}, 
 \label{ex1} 
\eeq
where $A$, $a$, $b$, are parameters and $\mathcal{P}_\zeta^{\text{SI}}(k)$ is the usual (nearly) scale invariant spectrum with spectral index $\ns-1\approx-0.04$ . We take $a = 3/2$ and $b = 1/2$ as an example. We plot this as the blue curve in Fig.~\ref{fig:constraint}. 
Here $\sigma^2 (M(k=1/R))$ is given by (ignoring the approximately scale invariant tails)
\beq
 &&\sigma^2 (M(1/R)) = A  \frac{16}{81} \lkk \sqrt{\pi}\, x\, (1 - 4 x^2 - 2 x^4) \right.  \nn
 &&~~~~~~~~~~~~~~~~ \left. + \pi\, x^4 (5 + 2 x^2) e^{x^2} (1 - {\rm erf} (x)) \rkk, 
\eeq
where $x = k_p R$. This function has a maximum $\sigma^2_{(\rm max)} \simeq 0.116 A$ at $x = 0.695$. We use this to compute the PBH fraction $\beta (M(k=1/R))$. We can then relate a constraint on the number of PBHs to a constraint on the amplitude of spike $A$ and in turn to a constraint on the power ${\cal P}_\zeta (k = k_p)$; this is shown as the black dashed line in Fig.~\ref{fig:constraint}. These constraints arise from extragalactic $\gamma$-rays from evaporation and consistency of big-bang nucleosynthesis~\cite{Carr:2009jm}, femtolensing of $\gamma$-ray bursts~\cite{Barnacka:2012bm}, white-dwarf explosions~\cite{Graham:2015apa}, microlensing~\cite{Griest:2013aaa, Novati:2013fxa, Tisserand:2006zx, Niikura:2017zjd}, and accretion effects~\cite{Ricotti:2007au}  (see Ref.~\cite{Carr:2016drx} for a review of these constraints). 

In Fig.~\ref{fig:constraint} we also plot other constraints on the amplitude of power spectrum. Note that we neglect some model dependent constraints that depend on the detailed nature of dark matter, and focus on model independent constraints. The observation of CMB temperature anisotropy gives a precise measurement of power spectrum at large scales. The large scale structure and Lyman-$\alpha$ forest observations put constraints on smaller scales. We adopted those constraints from Ref.~\cite{Bringmann:2011ut} and plot them as the purple line. On smaller scales, the $\mu$ and $y$-distortions are generated from the Silk damping of the perturbations. The COBE/FIRAS experiment puts the constraint on the amount of these distortions as $\mu \lesssim 9 \times 10^{-5}$ and $y \lesssim 1.5 \times 10^{-5}$~\cite{Fixsen:1996nj}, which can places an upper bound on the amplitude of power spectrum~\cite{Chluba:2012we}; we assume a delta-function power spectrum to plot this constraint. If the scalar perturbation are quite large, then second-order effects lead to the generation of gravitational waves~\cite{Saito:2008jc, Saito:2009jt}. The constraint on the energy density of gravitational waves by the pulsar-timing experiment, such as EPTA~\cite{Lentati:2015qwp}, PPTA~\cite{Shannon:2015ect}, and NANOGrav~\cite{Arzoumanian:2015liz}, can be recast into that of the amplitude of scalar perturbation. To plot the constraint, we adopt the calculation in Ref.~\cite{Inomata:2016rbd} assuming a delta-function power spectrum. 

In Ref.~\cite{Inomata:2017okj}, the authors claimed that the PBHs can be all the dark matter for masses of order $10^{20} \, {\rm g}$. In this case, the amplitude of power spectrum should be around $10^{-2}$ at $k \simeq 7 \times 10^{12} \ {\rm Mpc}^{-1}$. An example of such a power spectrum is shown in Fig.~\ref{fig:constraint}, where we assume \eq{ex1} with $A \sim 0.01$ and $k_0 \sim 7 \times 10^{12} \, {\rm Mpc}^{-1}$ (but with small deviations from these values for later convenience). Note that a relatively sharp peak in the power spectrum may be possible at around $k\sim 10^5\,{\rm Mpc}^{-1}$. Interestingly, this corresponds to black holes within an order of magnitude of $M \sim 30 M_\odot$, which is the range currently observed by LIGO/Virgo. Some analyses suggest that this might not be a significant fraction of the dark matter, though a small fraction is possible.

In the next section, we outline a method to reconstruct the inflaton potential from these types of power spectra. In Section \ref{sec:polynomial}, we provide a polynomial potential model that leads to similar power spectra.

\section{Reconstruction of inflaton potential}
\label{sec:formula}

We study the standard two-derivative action for gravity with a  single scalar field $\phi$. Without loss of generality, one can perform field re-definitions to obtain the familiar Einstein-Hilbert action, minimally coupled to $\phi$  
\beq
 S = \int d^4 x \sqrt{-g} \lkk {1\over 16\pi G}\mathcal{R} + \frac{1}{2} \del_\mu \phi \del^\mu \phi - U(\phi) \rkk. \,\,
\eeq
From here on we will work in natural units $\hbar=c=8\pi G=1$.

As we discussed in the previous section, we are interested in the case where the power spectrum has a peak at a cosmologically small scale. In this section, we describe a method to reconstruct the inflaton potential $U(\phi)$ from a given power spectrum, assuming $\epsilon$ ($\equiv (U'/U)^2/2$) $\ll1$, but $\eta$ ($\equiv U'' / U)$ can be larger than unity during inflation.\footnote{Reconstruction of full inflaton potential was discussed in Ref.~\cite{Hodges:1990bf}, where they assume the slow-roll approximation. Their method was generalized in Ref.~\cite{Copeland:1993jj}, which is similar to our method. 
On the other hand, reconstruction of ``local'' inflaton potential 
has been extensively studied, where some derivatives or slow-roll parameters 
are reconstructed from the spectral index and its derivatives. See, e.g., Refs.~\cite{Easther:1993qg, Liddle:2003py, Peiris:2008be, Aslanyan:2015hmi}. 
}

\subsection{Background Evolution}

Let us first describe the evolution of the FRW homogeneous background, which we take to be spatially flat. It is governed by the following equations: 
\beq
 &&N_e (t) = \int_{t_{\rm end}}^{t} H (t)\, d t ,
 \label{Ne-t}
\\
 &&\ddot{\phi} + 3 H \dot{\phi} + \frac{d U}{d \phi} = 0 ,
 \label{EOM}
 \\
 &&H^2 = \frac{1}{3} \lmk \frac{1}{2} \dot{\phi}^2 + U (\phi) \rmk, 
 \label{Friedmann}
\eeq
where $t_{\rm end}$ is the time at which inflation ends. Here and hereafter, dots represent derivatives with respect to time. The latter two equations can be rewritten exactly as 
\bea
&& \dot{H} = - \frac{\dot{\phi}^2}{2}, 
 \label{phi-t} \\
&& U = 3\, H^2 + \dot{H}. 
 \label{V-t}
\eea
An accelerating expansion, inflation, occurs when the following slow-roll parameter is much smaller than unity: 
\beq
 \epsilon_H \equiv 
 - \frac{\dd H}{\dd N_e} 
 = - \frac{\dot{H}}{H^2}. 
 \label{epsilon_H}
\eeq

\subsection{Fluctuations}

Fluctuations in scalar modes around the homogeneous background are described by the curvature perturbation $\hat{\zeta}$. Its variance in fluctuations are measured by the matter power spectrum: $\la \hat{\zeta} (k) \hat{\zeta} (k') \ra = (2\pi)^3 \delta^3 (k + k') (2 \pi^2 /k^3) {\cal P}_\zeta (k)$, where 
\beq 
 \mathcal{P_\zeta} (k)  
 = \frac{2 k^3}{8 \pi^2} \abs{\frac{v_{\bm k}}{a \sqrt{\epsilon_H}} }^2. 
 \label{P1}
\eeq
Here $k$ ($\equiv \abs{\bm k}$) is a comoving wavenumber of a Fourier mode and $v_{\bm k}$ is the mode function associated with the Mukhanov-Sasaki (MS) variable. The mode function obeys the MS equation~\cite{Sasaki:1986hm, Mukhanov:1988jd}
\beq
 \frac{\dd^2 v_{\bm k}}{\dd \eta^2} + \lmk k^2 - \frac{1}{z} \frac{\dd^2 z}{\dd \eta^2} \rmk v_{\bm k} = 0, 
\eeq
where $z \equiv a \sqrt{\epsilon_H}$ and $\eta$ is conformal time ($\dd \eta \equiv \dd t / a$). The initial condition for the quantum state is taken to be the standard Bunch-Davies vacuum, wherein
\beq
 v_{\bm k} \to \frac{1}{\sqrt{2k}} e^{-i k \eta}, \,\,\,\,\,\,\mbox{as}\,\,\,{k \over a H} \to \infty. 
\eeq
For the purpose of numerical simulation, it is convenient to rewrite the MS equation in terms of $\zeta_{\bm k}\equiv v_{\bm k}  / (a \sqrt{2 \epsilon_H})$: 
\beq
 \frac{\dd^2 \zeta_{\bm k}}{\dd \eta^2} 
 + \lmk 2 + \epsilon_{H2} \rmk aH \frac{\dd \zeta_{\bm k}}{\dd \eta}
 + k^2 \zeta_{\bm k} = 0, 
 \label{MSeq}
\eeq
where 
\beq
 \epsilon_{H2} \equiv 
 \frac{\dd \epsilon_H}{\dd N_e} 
 = \frac{\dot{\epsilon}_H}{H \epsilon_H}. 
\eeq

An infinitesimal change of comoving wavenumber between $k$ and $k+dk$ that corresponds to the scale leaving the horizon between time $t$ and $t+dt$ can be written as 
\beq
 d \ln k = d ( \ln (a H))  \simeq - d N_e = - H d t, 
 \label{k-Ne}
\eeq
where we use $\epsilon_H \ll 1$. This gives 
\beq
 \ln (k_{\rm end}/ k ) \simeq N_e,
\eeq
where $k_{\rm end}$ is the scale that leaves the horizon at $t =t_{\rm end}$.

\subsection{Key Approximation}

Now our key simplifying approximation is to neglect $\epsilon_{H2}$ in Eq.~(\ref{MSeq}) when we compute the power. We shall see that this is not precise in the non-slow-roll regime, where the potential term in the background equation of motion $U'(\phi)$ is sub-dominant to the acceleration $\ddot\phi$ and friction $3H\dot\phi$ terms. But in the regular slow-roll regime (where it is the acceleration term $\ddot\phi$ that is negligible), this assumption is precise. In any case, we shall take the power spectrum of the form
\beq 
 \Ps (k) \simeq \frac{H^2}{8 \pi^2 \epsilon_H} . 
 \label{P2}
\eeq
We shall use this formula to reconstruct the inflaton potential. After the reconstruction, we will numerically compare to the exact result from solving the MS equations exactly, finding that while there are corrections, it does not change the result tremendously and certainly suffices to capture the qualitative behavior.\footnote{
If $\epsilon$ decreases faster than about $a^{-3}$ during the non-slow-roll regime, 
the so-called decaying mode of $\zeta_{\bm k}$ grows faster on super-horizon scales than the constant mode that we used to derive \eq{P2}. 
Our exact result from solving the MS equations takes into account both modes, 
so that we will see that neglecting the decaying mode does not affect the qualitative behavior in the case we are interested in. 
}

Note that we will still use the {\em full} expression for $\epsilon_H=-\dot{H}/H^2$, rather than the simple $\epsilon=(U'/U)^2/2$ that is used in the usual slow-roll treatments. In particular, this means we do not demand that the acceleration $\ddot\phi$ is always negligible to $U'$; this is important near a critical point in the potential where $U'\to0$. So we are still going beyond the standard slow-roll regime and capturing, at least qualitatively, the non-slow-roll regime too.

\subsection{Reconstruction} 

Now we would like to derive formulas that allow us to calculate the potential for a given $\Ps(k)$. First one comes from Eqs.~(\ref{epsilon_H}), (\ref{P2}), and (\ref{k-Ne}): 
\beq
 \frac{1}{H^4} \frac{d H^2}{ d \ln k} =  \frac{1}{4 \pi^2 \Ps}.
 \label{k-U2}
\eeq 
So we can integrate $d H^2$ as 
\beq
 \frac{1}{H^2(k)} - \frac{1}{H_*^2} = - \frac{1}{4 \pi^2} \int_{k_*}^k \frac{1}{ \Ps(k')} d \ln k'.
 \label{formula3}
\eeq
Here we defined an arbitrary pivot scale $k_*$ and corresponding Hubble parameter $H_*$. The potential $U$ as a function of $k$ can be calculated from Eq.~(\ref{V-t}) as
\beq
 U (k) = 3 H^2 - \frac{H^4}{8 \pi^2 \Ps(k)}. 
 \label{V-k2}
\eeq
Although the second term in the right hand side is much smaller than the first one in the regular slow-roll regime $\epsilon \ll 1$, it is important in the non-slow-roll regime as we explain below. 

The potential can be implicitly determined by the power spectrum, by obtaining the inverse function of $\phi = \phi(k)$. This can be determined by first using Eq.~(\ref{phi-t}) to express $\dot\phi$ as
\beq
\frac{d \phi}{d t} = \sqrt{\frac{d H^2}{d \ln k}} .
\eeq
Then we use Eq.~(\ref{k-U2}) to obtain
\beq
 \phi (k) - \phi_* = - \int_{k_*}^k \sqrt{\frac{H^2}{4 \pi^2 \mathcal{P_R}}}  d \ln k'. 
 \label{formula4}
\eeq
Note that we can always shift the field value such that $\phi_*= 0$. Thus there is only one unknown parameter $H_*$ in the above formulas, which defines a height of the potential at the pivot scale $U_*$. By using \eq{V-k2} and an inverse function of (\ref{formula4}), we can calculate a one-parameter family of potentials $U$ that reproduces exactly the same spectrum. 

The shape of the spectrum is conveniently specified by the spectral index $\ns(k)$ as 
\beq
 \ns(k) - 1 \equiv \frac{d \ln \Ps}{d \ln k}. 
\eeq
The spectral index can be written by slow-roll parameters 
as 
\beq
\ns (k) -1  = 2 \epsilon_H (k) + \epsilon_{H2} (k). 
\label{nsgen}\eeq
Note that in the regular slow-roll regime $\epsilon_{H}$, $\epsilon_{H2} \ll 1$, the usual slow-roll parameters $\epsilon$ and $\eta$ can be written as $\epsilon \simeq \epsilon_H$, $\eta \simeq \epsilon_{H2}/2 + 4 \epsilon_H$, which reproduces the familiar $\ns\simeq 1-6\epsilon+2\eta$. But in general and in the non-slow-roll regime, when $\epsilon_{H2}$ is not necessarily small, the more general expression of (\ref{nsgen}) is required.

We can write the slow-roll parameters $\epsilon_H$ and $\epsilon_{H2}$ in terms of the spectrum $\Ps$ and potential $U$ as. 
\bea
&&\epsilon_H (k)  = \frac{H^2 (k) }{8 \pi^2 \Ps(k)}, \\ 
&&\epsilon_{H2} (k)  =  - 2 \epsilon_{H2}  + \frac{d \ln \mathcal{P_R}}{d \ln k}. 
 \label{condition}
\eeq
This shows that $\epsilon_{H2} \gtrsim 1$ when $d \ln \mathcal{P_R} / d \ln k \gtrsim 1$. Thus we need to consider the non-slow-roll inflation to generate a large amplitude of power spectrum at small scales~\cite{Motohashi:2017kbs}. 

Finally, we comment on the second term in \eq{V-k2}. One might think that the second term is negligible for $\epsilon_H \ll 1$.  However, the second term is relevant for the equation of motion, in which the derivative of potential is important. 
To see this, let us calculate 
\beq
 \frac{d U}{d \phi} 
 = \frac{d t}{d \phi} \frac{d}{dt} \lkk \lmk 3 - \epsilon_H \rmk H^2 \rkk. 
\eeq
The derivatives are given by 
\bea
&& 3 \frac{d}{dt} H^2 = -6 \epsilon_H H^3 , \\
&& H^2 \frac{d}{dt} \epsilon_H = \epsilon_H \epsilon_{H2} H^3, 
\eea
which shows that the second term is as large as the first term when $\epsilon_{H2}$ is of order unity or larger, as it is in the non-slow-roll inflation regime.

\section{Examples}
\label{sec:applications}

In this section we carry out the above program of reconstructing the inflaton potential from a given power spectrum by using the above formulas derived. We also solve the MS equation (\ref{MSeq}) numerically and check that the power spectrum calculated from the reconstructed potential is qualitatively in agreement with the original one. 

\subsection{Analytically Tractable Models} 

We first give two simple examples where calculations can be done analytically, before moving to more complicated examples which will required numerics. 

\subsubsection{Flat Power Spectrum} 

Let us begin by considering the special case in which the spectrum is perfectly flat, i.e., $\Ps (k) = A = {\rm const}$. From Eq.~(\ref{formula3}), we obtain 
\beq
 H^2(k) = \lkk \frac{1}{H^2_*} - \frac{1}{4 \pi^2 A} \ln \frac{k}{k_*} \rkk^{-1}. 
\label{Hex1}\eeq
Inserting this into Eq.~(\ref{formula4}) and carrying out the integral we obtain
\be
 \phi(k) = \phi_* - 2 \sqrt{\frac{4 \pi^2 A }{ H_*^2} - \ln \frac{k}{k_*}}  + 2 \sqrt{\frac{4 \pi^2 A }{H_*^2}}, 
 \label{phisol1}
\ee
Inverting this for $k=k(\phi)$ and inserting into the expression for the Hubble parameter (\ref{Hex1}) we obtain
\beq
 H^2 (k (\phi)) = \frac{16 \pi^2 A}{\lmk \phi - \phi_* - \sqrt{\frac{16 \pi^2 A }{ H_*^2}} \rmk^{2}},
\label{Usol1}
\eeq
with the corresponding potential from Eq.~(\ref{V-k2}) given by
\beq
 U(\phi) = 3 H^2 (\phi) - \frac{H^4 (\phi)}{8 \pi^2 A}. 
\eeq
We have numerically checked that by solving for the perturbations in this (moderately complicated) potential we indeed get an approximately flat spectrum.

\subsubsection{Flat Power Spectrum with a Discontinuity} 
\label{sec:analytic2}

Next we consider the case where the spectrum has a discontinuity, suddenly changing from one flat spectrum with amplitude $A_1$ to another flat spectrum with amplitude $A_2$ at a wavenumber $k_1$ as
\beq
 &&\Ps (k) = A_1  \qquad \text{for }  k_* < k < k_1,  \nonumber \\
 &&\Ps (k) = A_2  \qquad \text{for }  k_1 < k. 
\eeq
In each regime, the solution can be written in the form of Eqs.~(\ref{phisol1}) and (\ref{Usol1}). Defining 
\beq
 &&H_1^2  = \lkk \frac{1}{H^2_*} - \frac{1}{4 \pi^2 A_1} \ln \frac{k_1}{k_*} \rkk^{-1}, \\
 &&\phi_1 =  \phi_* - 2 \sqrt{\frac{4 \pi^2 A_1 }{ H_*^2 }- \ln \frac{k_1}{k_*}} + 2 \sqrt{\frac{4 \pi^2 A_1 }{ H_*^2}}, 
\eeq
we can write the solution as 
\beq
 &&H^2 (k(\phi)) = \frac{16 \pi^2 A_1}{\lmk \phi - \phi_* - \sqrt{\frac{16 \pi^2 A_1 }{ H_*^2}} \rmk^{2}} 
 \quad \text{for }  k_* < k < k_1,  \nonumber\\
 &&H^2 (k (\phi)) = \frac{16 \pi^2 A_2}{\lmk \phi - \phi_1 - \sqrt{\frac{16 \pi^2 A_2 }{ H_1^2}} \rmk^{2}} 
 \quad \text{for }  k_1 < k. 
\eeq
The potential $U(\phi)$ is then calculated from \eq{V-k2}. 

Note that the potential itself has a discontinuity at $\phi = \phi_1$, just as the power spectrum does. Interestingly, in the case of $A_1 < A_2$, the potential jumps to a higher value as $\phi$ crosses $\phi_1$. We now check that the classical evolution of the inflaton can actually climb this potential barrier at $\phi = \phi_1$. The kinetic energy of the inflaton before it reaches $\phi_1$ is given by 
\beq
 {\dot{\phi}^2\over2}  = - \dot{H}  = \epsilon_H\, H^2  = \frac{H^4}{8 \pi^2 A_1}. 
 \label{kineticE}
\eeq
The difference in the potential energy across the discontinuity is given by 
\beq
 \Delta V 
 &=& {\rm lim}_{\delta \to 0} \lmk U( \phi_1 + \delta) - U(\phi_1 - \delta)  \rmk ,
 \nonumber\\
 &=& \frac{H^4}{8 \pi^2} \lmk \frac{1}{A_1} - \frac{1}{A_2} \rmk. 
\eeq
This is always smaller than the kinetic energy (\ref{kineticE}), so that the inflaton can in fact climb the potential at $\phi = \phi_1$.

\subsection{Numerical Calculations} 

Here we would like to consider more complicated models which are no longer analytically tractable. In this case we need to  calculate Eqs.~(\ref{formula3}) and (\ref{formula4}) numerically for some input power spectrum and then insert into Eq.~(\ref{V-k2}) to obtain the reconstructed potential.

\subsubsection{Flat Spectrum with a Rapid Change}

We use the following toy model to improve on the model of Section \ref{sec:analytic2}, where the discontinuity in the power spectrum  from values $A_1$ to $A_2$ at some scale $k_1$ is now smoothed out. As a useful example, we use a hyperbolic tangent as the smoothing function described by some smoothing scale $\sigma_k$: 
\beq
 \Ps (k) = A_1 + (A_2 - A_1) \lkk {\rm tanh} \lmk \frac{\ln (k/ k_1)}{\sigma_k} \rmk - 1 \rkk. \,\,\,
 \label{Ex3}
\eeq
We take $A_1 = 0.1$, $A_2 = 1$, $\sigma_k = 1$, and $k_1 = 10^{13}$ to illustrate the results (of course these are not realistic values for cosmology, but we return to realistic parameters in the next Section). 

\begin{figure}[t] 
   \centering
   \includegraphics[width=8.5cm,height=6.5cm]{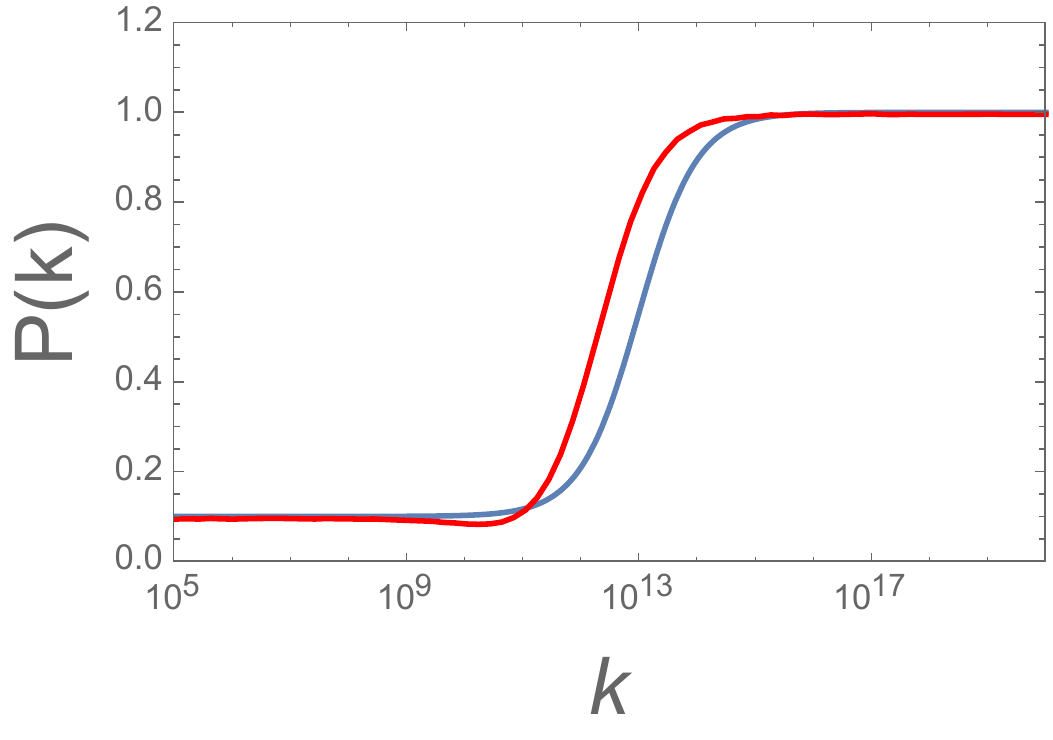} 
   \caption{Example of an input hyperbolic tangent power spectrum (blue curve). We reconstruct a potential from this power spectrum making use of the approximation \eq{P2}. To check the consistency of our method, we numerically calculate a power spectrum (red curve) from the reconstructed potential without any approximation. } 
   \label{fig:dis}
\end{figure}

We first reconstruct the potential by using the method of Section \ref{sec:formula}. By comparison, we also solve the MS equation (\ref{MSeq}) 
and calculate the power spectrum from the reconstructed potential to check the accuracy of our method. The comparison is shown in Fig.~\ref{fig:dis}. Here the blue line is the input power spectrum (\ref{Ex3}) and the red line is the one calculated from \eq{P2} using the reconstructed potential. 
We can see that the approximate spectrum is in good qualitative agreement with the exact one. 

\subsubsection{Spectrum with a Spike at a Small Scale}
\label{sec:app2}

Next we move to more interesting and realistic power spectra. Here we input the power spectrum of Eq.~(\ref{ex1}); it is shown as the solid blue line in Fig.~\ref{fig:constraint}. This spectrum is physically motivated by having significant PBH formation due to a spike in the spectrum and it is consistent with existing observations.  

\begin{figure}[t] 
   \centering
\hspace{-0.5cm}  \includegraphics[width=9cm,height=7cm]{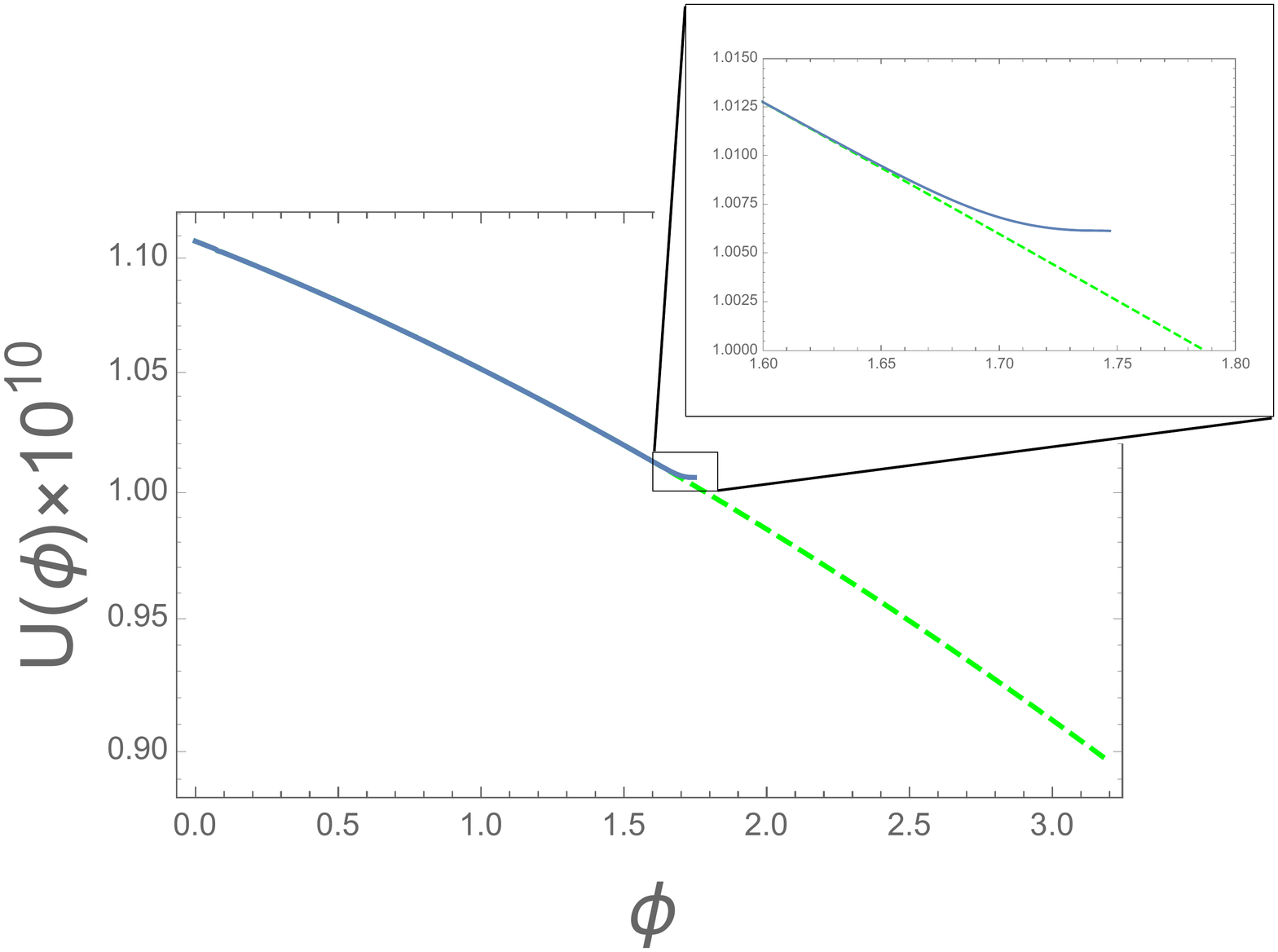} 
   \caption{Reconstructed potential from the power spectrum of Eq.~(\ref{ex1}), which is representative of a spectrum with a large spike in it on small scales and is shown as the blue curve in Fig.~\ref{fig:constraint}. Here we assume $U_* = 24 \pi^2 {\cal P_R} (k_*) /10^3$. We plot the case with the spike (blue) and without the spike (green dashed).}
   \label{fig:reconstruct}
\end{figure}

A reconstructed potential from spectrum of Eq.~(\ref{ex1}) is shown in Fig.~\ref{fig:reconstruct}, where we take the overall height of the potential to be $U_* = 24 \pi^2 \Ps (k_*) /10^3$. We do not plot the potential for $\phi$ larger than about $1.75$, which corresponds to the field value at which a mode at the right edge of Fig.~\ref{fig:constraint} ($k = 10^{18} \ {\rm Mpc}^{-1}$) crosses the horizon. We also plot a reconstructed potential from a power spectrum without the spike (a featureless potential); this is the dashed green line. The field value reaches about $3.2$ when the mode $k = 10^{18} \ {\rm Mpc}^{-1}$ crosses the horizon. We zoom in on a region where the reconstructed potential deviates somewhat from the featureless one. This shows that a small difference of inflaton potential can result in a large deviation for the power spectrum, especially when one curve has a derivative that is becoming small at one point when the other is not. 

An interesting observation is that PBH formation from primordial density perturbations might not be consistent with some constraints unless the spike is rather narrow. This in turn requires an ordinary slow-roll condition to be violated $\abs{\eta} \ll 1$ ~\cite{Motohashi:2017kbs}. To generate significant numbers of PBHs, we need $\Ps \sim 10^{-2}$. Then the slope of the spectrum 
has to be appreciable $d \ln \mathcal{P_R} / d \ln k \gtrsim 1$ in some range in the interval of $(10 \, {\rm Mpc}^{-1}, 10^{13} \, {\rm Mpc}^{-1})$. This means that $\epsilon_{H2}$ and $\eta$ are as large as of order unity and we cannot use the usual slow-roll approximation in these scales if one demands high accuracy in computation. 

We have checked that the approximate formula (\ref{P2}), which neglects this correction, is nonetheless a reasonable approximation in this example. We solved the MS equation with the reconstructed potential and obtain the resulting power spectrum. The result is shown as the red line in Fig.~\ref{fig:constraint}. We can see that it is qualitatively in agreement with the original one.

\section{PBH formation and Polynomial potential} 
\label{sec:polynomial}

According to the results of the previous section, we need to violate the usual slow-roll conditions in which $\ddot\phi$ is always subdominant to $U'$ to generate PBHs from primordial perturbation, and enter the non-slow-roll regime. We also found that the inflaton potential can have a positive gradient during this non-slow-roll regime~\cite{Germani:2017bcs, Ballesteros:2017fsr}. In this section, we provide an example of a simple polynomial potential, with a small local minimum and corresponding small hill at which a large perturbation is generated. This is found to produce PBHs and satisfies all known constraints. We then remark on the fine-tuning required to achieve this.

\subsection{Polynomial Potential}

Motivated by an effective theory with a cutoff $\Lambda$, we consider the following polynomial potential: 
\beq
 U(\phi)  =  U_0 
 \lkk
 1 + c_1 \frac{\phi}{\Lambda} 
 + \frac{c_2}{2} \frac{\phi^2}{ \Lambda^2}
 + \frac{c_3}{3!} \frac{\phi^3}{ \Lambda^3}
 + \frac{c_4}{4!} \frac{\phi^4}{ \Lambda^4}
 + \frac{c_5}{5!} \frac{\phi^5}{ \Lambda^5} 
 \rkk, 
 \nonumber\\
\label{Uquintic}\eeq
where $U_0$ is the energy scale of inflaton potential and $c_i$ are constants. In principle one can (and should) add a tower of higher order terms, but we truncate the potential here to this quintic polynomial, as it suffices to illustrate our main points.

We can shift and redefine the origin of $\phi$ such that $c_2 = 0$ at $\phi = 0$ without loss of generality. We can also set $c_4 = 1$ by rescaling $\Lambda$ without loss of generality. We consider the case where (slow-roll) inflation starts and the CMB scale leaves the horizon when $\phi$ is in the vicinity of the origin, where we can neglect terms higher order than cubic
\beq
 U(\phi) \simeq U_0 \lkk 1 + c_1 \frac{\phi}{\Lambda} + \frac{c_3}{3!} \frac{\phi^3}{\Lambda^3} 
 \rkk. 
\label{Ucubic}\eeq
The inflaton starts to roll fast after the time at which $|U'' / U| \simeq 1$ or $\phi \simeq \Lambda^3 / c_3$. We denote the number of e-foldings number after the ordinary slow-roll phase as $N_{\rm NSR}$, as this enters the non-slow-roll phase. We choose parameters such that $\epsilon_{H}$ is much smaller than unity and so inflation continues even after the ordinary slow-roll phase ends. We are interested in the case where $\epsilon_H$ decreases by a factor of order $10^{-7}$ during this regime so that PBHs can be later formed from these primordial fluctuations. 

The maximum number of e-foldings that can be realized in the slow-roll region ($|\phi| \lesssim \Lambda^3 / c_3$) is given by 
\beq
 N_{\rm max,0} \simeq \sqrt{2}\, \pi \frac{\Lambda^2}{\sqrt{c_1 c_3}}, 
\eeq
where we assume $U(\phi) \simeq U_0$. The spectral index on large scales corresponding to the CMB ($N_e = N_{\rm CMB} = 50-60$) can be computed using the standard slow-roll methods giving~\cite{Baumann:2007ah, Linde:2007jn}
\beq
 1-n_s \simeq \frac{4 \pi}{N_{\rm max,0}} {\rm cot}(x),
 \label{ns}
 \eeq
where
 \beq
x \equiv  \pi \frac{N_{\rm CMB} - N_{\rm NSR}}{N_{\rm max,0}}, 
\eeq
and we assume $N_{\rm max,0} \gg 1$. We determine $c_1$ so that the resulting spectral index is consistent with the measured spectrum on the CMB by the Planck satellite (and turns out to be $c_1\sim -10^{-4}$). The energy scale of inflaton potential $U_0$ is determined such that the amplitude of curvature perturbation at the CMB scale is consistent with the COBE normalization. This requires 
\beq 
 {\cal P_\zeta} (k_*) \simeq \frac{N_{\rm CMB}^4\, c_3^2\, U_0}{48 \pi \Lambda^6} 
 \frac{\sin^4 (x)}{x^4}. 
\eeq

The higher order terms in (\ref{Uquintic}) that take us beyond the cubic approximation in (\ref{Ucubic}) are important as $\phi$ increases out of the slow-roll (CMB) stage and enters the non-slow-roll phase. The parameter $c_5$ is fine-tuned so that the velocity of inflaton and $\epsilon_H$ can decrease significantly near a small hill in the potential and yet still make it over the top to the other side. 
We find that significant fine-tuning is required in $c_5$ so that the amplitude of power spectrum can be as large as $10^{-2}$; which we will analyze further in Section \ref{sec:finetuning}. In the examples we use in this paper, $c_5$ is taken to be a certain value 
about $0.55-0.6$ with a precision of order $10^{-7}\%$. 

As we see below, the parameter $c_3$ determines the e-folding number realized near the hilltop. In this paper, we take $c_3 = 0.5-0.65$ as an example. We check that the result does not change qualitatively even if we change $\Lambda$ within the range of $(0.2, 0.4)$, though we take $\Lambda = 0.3$ (in reduced Planck units) for definiteness. 

\begin{figure}[t] 
   \centering
   \includegraphics[width=8cm,height=7cm]{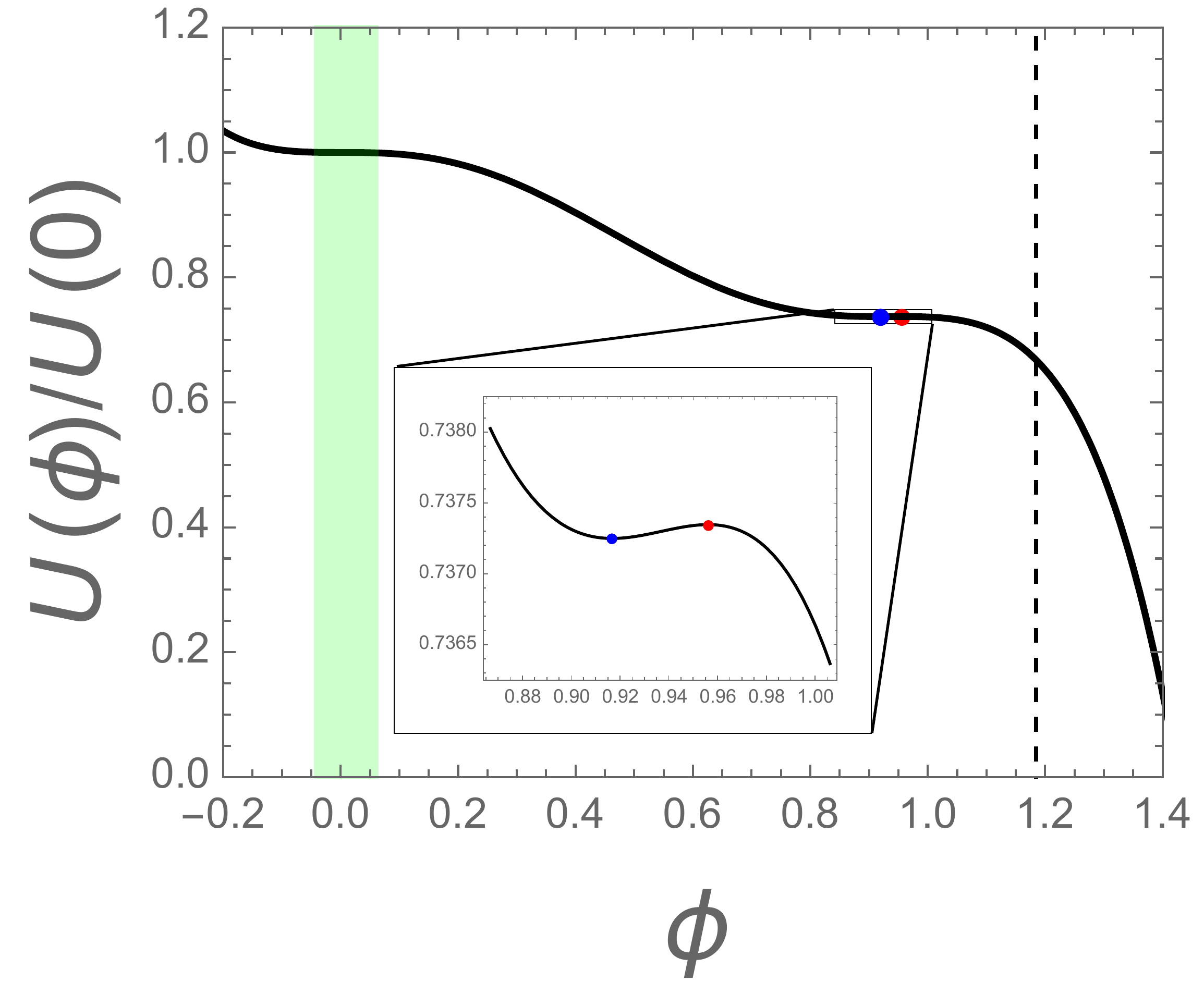} 
 \\  \vspace{0.5cm}
 \hspace{-0.5cm}  \includegraphics[width=8.2cm,height=7cm]{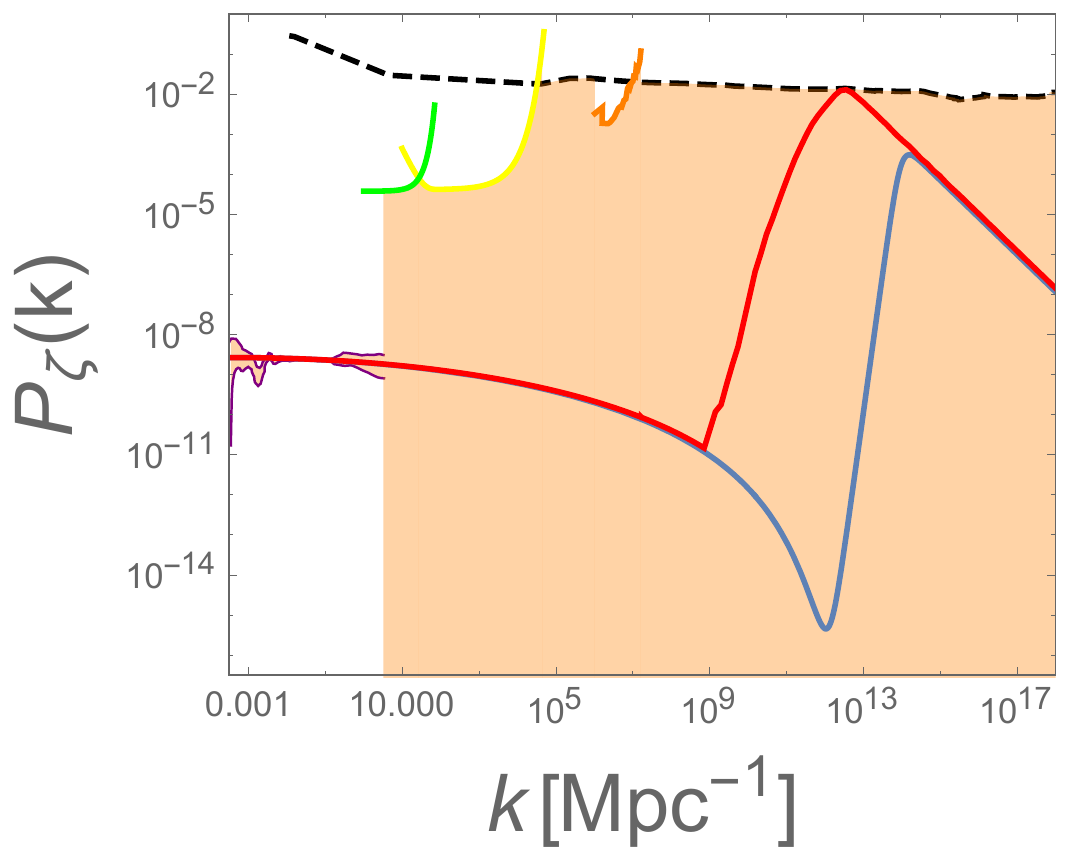} 
   \caption{ Upper panel: Quintic polynomial potential that gives rise to PBHs. The light-green colored region represents the domain of $\phi$ where the standard slow-roll approximation is accurate. The blue (red) dot represents a local minimum (maximum). Inflation ends when the inflaton reaches the black-dashed line. We take $N_{\rm max,0} = 65$, $c_3 = 0.52$, $c_5 = - 0.64072504$, and $\Lambda = 0.3$. Lower panel: Corresponding power spectrum. The red curve is the exact result from solving the MS equation and the blue curve is an approximate result using Eq.~(\ref{P2}).}
   \label{fig:polynomial}
\end{figure}

An example of a potential is shown in the upper panel of Fig.~\ref{fig:polynomial}, where we take $N_{\rm max,0} = 65$, $c_3 = 0.52$, $c_5 = -0.64072504$, and $\Lambda = 0.3$. The light-green colored region represents the domain of $\phi$ where we can use the slow-roll approximation is accurate. The blue (red) dot represents a local minimum (maximum) where one can obviously {\em not} assume that $\ddot\phi$ is negligible compared to $U'$ as $U'\to0$ at the minimum. It is the acceleration term that allows the field to roll over the small hill; see the inset for a close-up of this shape. Finally, inflation terminates when the inflaton reaches the black-dashed line. 

\subsection{Results}

We solve the equation of motion of inflaton \eq{EOM} with the Friedmann equation \eq{Friedmann}. We then solve the MS equation from the Bunch-Davis vacuum and calculate the power spectrum from \eq{P1}. The resulting power spectrum is shown as a red line in lower panel of Fig.~\ref{fig:polynomial}. We also calculate the power spectrum using the approximation \eq{P2} and the result is shown as the solid blue line. These results are rather different, though they do share some qualitative similarities. We see that the full result by solving \eq{MSeq} exactly is important here for accuracy.

\begin{figure}[t] 
   \centering
   \includegraphics[width=8cm,height=7cm]{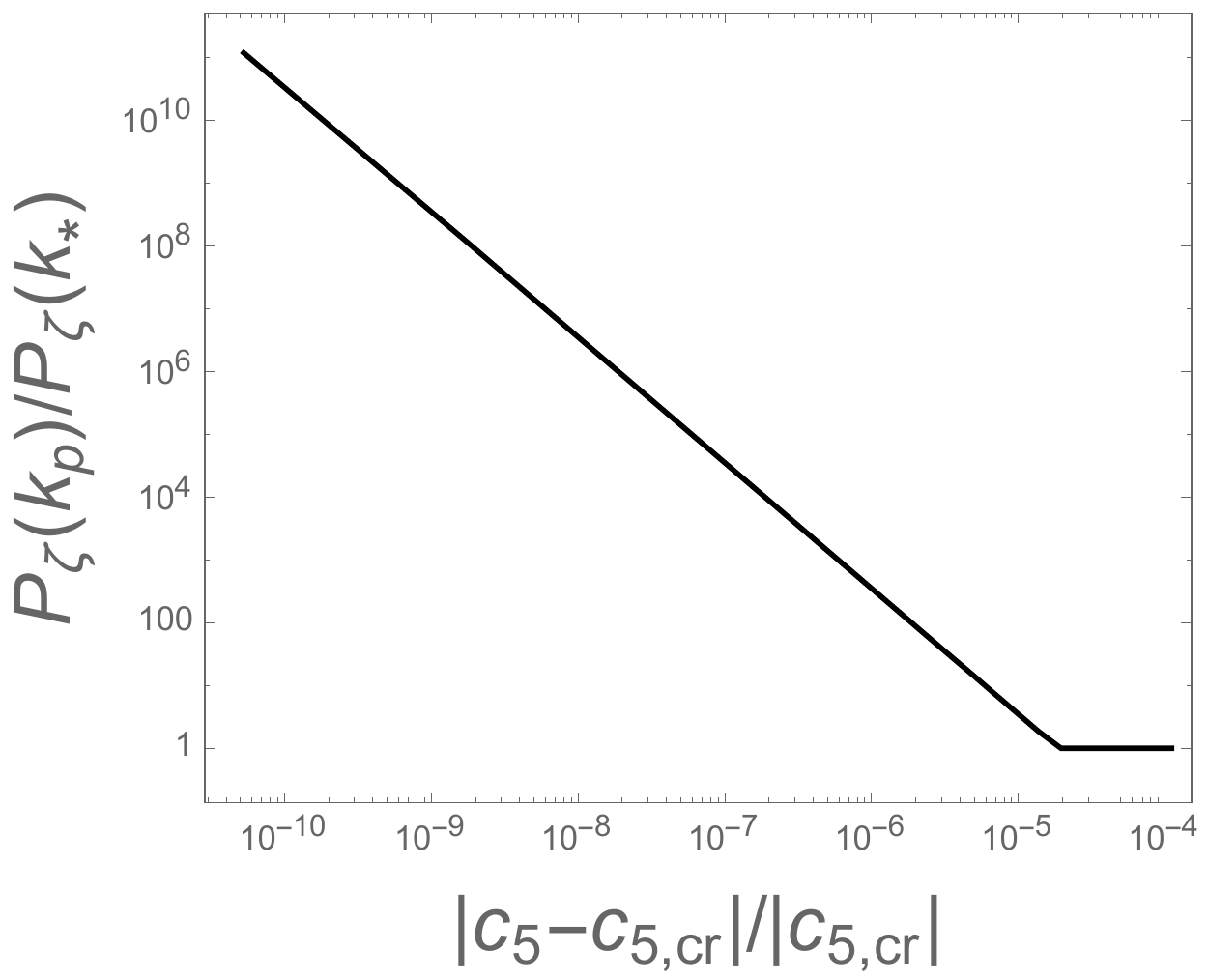} 
   \\ \vspace{0.5cm}
   \includegraphics[width=8cm,height=7cm]{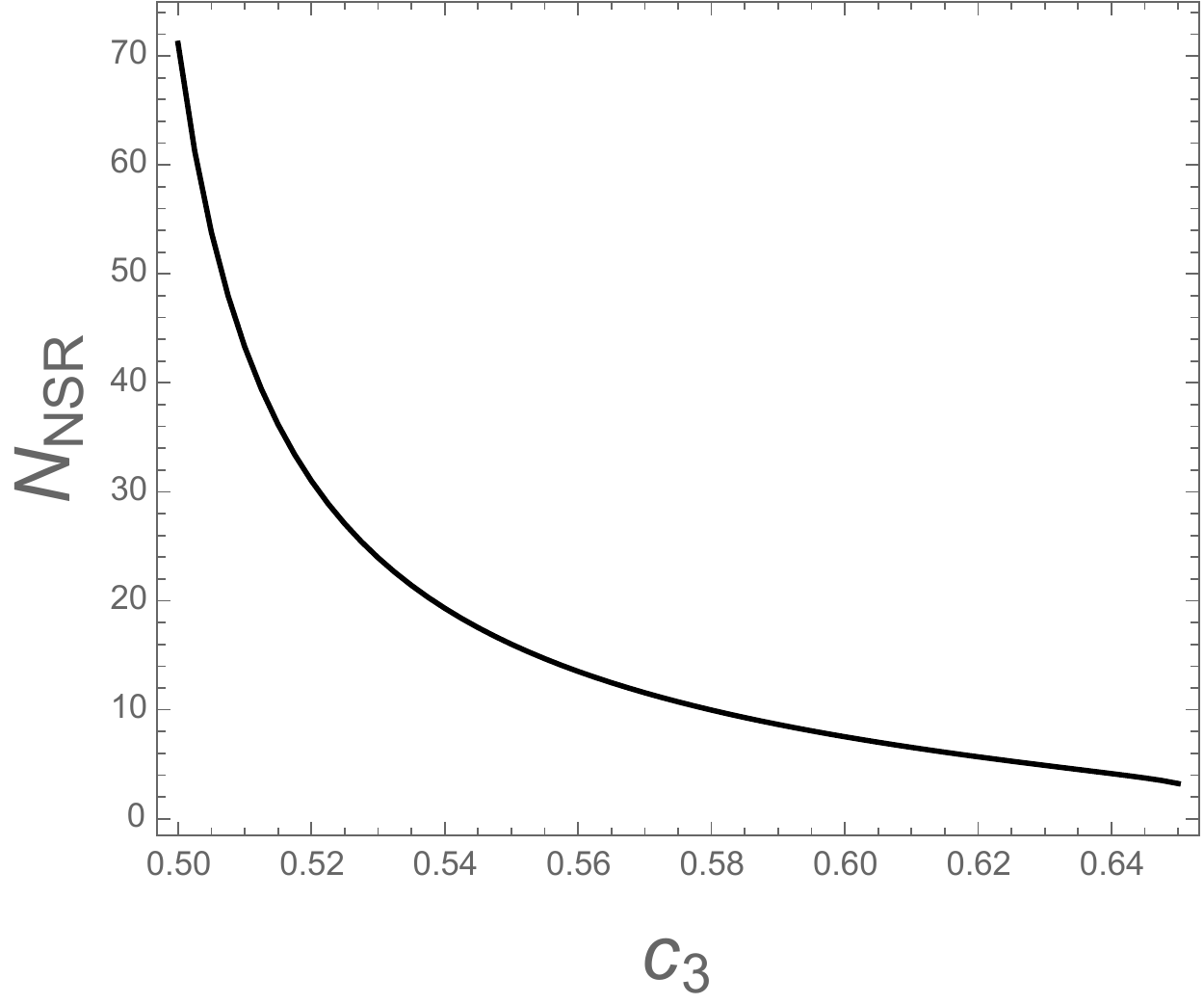} 
   \caption{Upper panel: Ratio of the amplitude of power spectrum at a maximum ($k = k_p$) and at a pivot scale $(k = k_*)$ as a function of the amount of $c_5$, with other parameters fixed to be $c_3 = 0.52$ and $\Lambda = 0.3$. 
   Lower panel: The number of e-foldings after the start of the non-slow-roll phase $N_{\rm NSR}$ ($= N_{\rm tot} - N_{\rm max, 0}$) as a function of $c_3$, with other parameters fixed to be $\abs{c_5 - c_{5,{\rm cr}}} / \abs{c_{5,{\rm cr}}} \simeq 10^{-8.5}$ and $\Lambda=0.3$. Note we solved the MS equation exactly in this figure.}
\label{fig:peak}
\end{figure}

We change the parameters $c_3$ and $c_5$ to see the parameter dependences of power spectrum. The upper panel of Fig.~\ref{fig:peak} shows the maximum amplitude of power spectrum as a function of of $c_5$, with all other parameters fixed (we used the exact numerical solution of the MS equation to produce Fig.~\ref{fig:peak}.) We use $\abs{c_5 - c_{5,{\rm cr}}} / \abs{c_{5,{\rm cr}}}$ to represent the amount of fine-tuning, where $c_{5,{\rm cr}}$ is the value of $c_5$ 
at which the power spectrum diverges at a small scale; this is associated with the critical shape of the potential $U$ where the classical evolution is unable to roll over the small hill. We can see that we need at least $\sim10^{-3}\%$ fine-tuning to realize an enhancement. The amplitude can be enhanced by an amount of order $10^7$ compared with the pivot scale when $c_5$ is fine-tuned by $\sim 10^{-6.5}\%$; we discuss this further in Section \ref{sec:finetuning}. 

The e-folding number realized after the end of slow-roll, $N_{\rm NSR}$, depends on $c_3$. We plot $N_{\rm NSR}$ ($= N_{\rm tot} - N_{\rm max,0}$) in the bottom panel of Fig.~\ref{fig:peak}, where $N_{\rm tot}$ is the total e-folding number from the beginning of inflation. We can see that $N_{\rm NSR}$ decreases as $c_3$ increases. Thus the parameter $c_3$ determines the e-folding number realized in the vicinity as $\phi$ rolls over the top of the small hill in the potential. This is important to calculate the total e-folding number and the spectral index from \eq{ns}. 

We checked that the difference in e-folding number between the one at the pivot scale and the one at the peak changes by $\approx1.5$ when $c_3$ changes from $0.5$ to $0.65$. This is quite small compared with $N_{\rm max,0}$ and $N_{\rm NSR}$. This implies that the inflaton reaches the top of the hill soon after the slow-roll ends. Since the power spectrum reaches the maximal value within a few e-folds after the end of slow-roll, we should take $N_{\rm CMB} - N_{\rm NSR} \simeq 30$ to generate PBHs with $M \sim 10^{20} \, {\rm g}$, and correspondingly even larger values of $N_{\rm NSR} $ for even larger PBHs. The observed spectral index $n_s^{(\rm obs)} \simeq 0.965$ can be realized if the maximum number of e-foldings is $N_{\rm max,0} \simeq 65$.

\subsection{Level of Fine-Tuning}
\label{sec:finetuning}

In this section we provide an explanation as to why $P_\zeta (k_p) / P_\zeta (k_*)$ roughly scales as $\propto (\abs{c_5 - c_{5,{\rm cr}}} / \abs{c_{5,{\rm cr}}} )^{-2}$, as seen in upper panel of Fig.~\ref{fig:peak}. Here, the critical value $c_{5,{\rm cr}}$ is defined by the value of $c_5$ above which the potential has a sufficiently deep local minimum causing the classical evolution of $\phi$ to become trapped there. {\em Note that we will only use very rough estimates in this subsection to capture the qualitative physics, rather than for acurracy.}

First, we note that the power spectrum can be very roughly estimated using the slow-roll approximation as
\beq
 \mathcal{P_\zeta} (k)  \sim \frac{H^4}{\dot{H}} \sim \frac{H^4}{\dot{\phi}^2}. 
 \label{P}
\eeq
This diverges as $\dot{\phi} \to 0$, which is realized as $c_5 \to c_{5,{\rm cr}}$. Suppose that $c_5$ is just below the critical value. Since we are interested in a peak of power spectrum $k_p$, we shall focus on a minimum value of $\dot{\phi}$ during the evolution of $\phi$. Since $\ddot{\phi} = 0$ at the minimum, the equation of motion implies $\dot{\phi} = - \frac{U'}{3 H}$ and we obtain 
\beq
 \mathcal{P_\zeta} (k_p)  \sim \frac{H^6}{U'^2}, 
\eeq
at a peak. Near the critical value, $c_5$ is close to $c_{5,{\rm cr}}$ and the time $t$ is close to $t_*$ the corresponding time at which $\dot\phi=0$ in the critical case. The derivative of the potential can be calculated perturbatively by a double Taylor series as 
\bea
&& U' (t_* + \delta t_*, c_{5,{\rm cr}} + \delta c_5)  \nonumber\\
&& \simeq  U' (t_*, c_{5,{\rm cr}})  +  \dot{\phi} U'' (t_*, c_{5,{\rm cr}}) \delta t_*  +  \frac{\del U'}{\del c_5} (t_*, c_{5,{\rm cr}}) \delta c_5. \,\,\,\,\,\,\,\,\,\,\,
\label{Taylor}\eea
where we have parameterized the distance from criticality by $t=t_* + \delta t_*$ and $c_5 = c_{5,{\rm cr}} + \delta c_5$. Now the first and second terms in (\ref{Taylor}) vanish because $U' \propto \dot{\phi} = 0$ at $t = t_*$ for $c_5 = c_{5,{\rm cr}}$ in the slow-roll approximation. The coefficient of the third term is just given by $U_0 \phi_*^4 / (4! \Lambda^5)$ and is nonzero. Thus we obtain the estimate
\beq
 \mathcal{P_\zeta} (k_p)  \sim \frac{U_0^3}{U_0^2 (\delta c_5)^2 } \frac{\Lambda^{10}}{\phi_*^8}, 
\eeq
where we used $H^6 \sim U_0^3$ for small $\dot{\phi}$. The power spectrum at the CMB scale is given by 
\beq
 \mathcal{P_\zeta} (k_*)  \sim \frac{U_0^3}{c_1^2 U_0^2} \Lambda^2. 
\eeq
Thus we obtain our very rough estimate for the spike in power
\beq
 \frac{\mathcal{P_\zeta} (k_p) }{\mathcal{P_\zeta} (k_*) } \sim \frac{c_1^2}{(\delta c_5)^2} \lmk \frac{\Lambda}{\phi_*} \rmk^8. 
\eeq
Now we are considering models with typical field displacements $\phi_* \sim \Lambda$ when this peak appears. Also $c_1$ is determined by the CMB normalization, which requires $c_1\sim -10^{-4}$. Hence we obtain
\beq
 \frac{\mathcal{P_\zeta} (k_p) }{\mathcal{P_\zeta} (k_*) } \simeq \mbox{Max}\left\{C \frac{10^{-8}}{(\delta c_5)^2}\,,\,1\right\}.
\eeq
where we included a fudge factor $C$ that comes from $(\Lambda / \phi_*)^8$ and other numerical factors that we have omitted, and also to capture the fact that these slow-roll estimates are inaccurate in this regime. This result is consistent with the precise result of Fig.~\ref{fig:peak} if we take $C \sim 0.01$. Since $\mathcal{P_\zeta} (k_*)\sim 10^{-9}$ and it is desirable to achieve $\mathcal{P_\zeta} (k_p) \sim 10^{-2}$ for significant PBH production, this requires $\delta c_5\sim 10^{-8.5}=10^{-6.5}\%$ fine-tuning.

\section{Conclusions and Discussion} 
\label{sec:conclusion}

PBHs are a fascinating possibility in modern cosmology, which could conceivably have some bearing on the dark matter, BH merger events observed by Advanced LIGO and Advanced Virgo, or other astrophysical phenomena such as acting as the seeds for supermassive BHs. Any of these possibilities is highly speculative and already somewhat constrained from existing data, but deserves to be fully investigated. A PBH requires very large over-densities to be produced in the early universe which could plausibly arise from some non-trivial behavior during inflation. In this paper we examined the canonical class of single field inflation models, organized by the standard two-derivative action, which is entirely specified then by the inflaton potential function $U(\phi)$.

We provided a method to reconstruct the inflaton potential $U(\phi)$ from a given power spectrum $\Ps(k)$, which is a useful tool on its own. But in particular, it was useful to show that the slow-roll approximation needs to be violated in order to obtain a sufficiently narrow spike in the spectrum, in agreement with other works in the literature~\cite{Garcia-Bellido:2017mdw, Ezquiaga:2017fvi, Kannike:2017bxn, Germani:2017bcs, Motohashi:2017kbs, Ballesteros:2017fsr}. This requires the inflaton to enter a regime where the potential is near a critical point. In this paper we exploited the idea of having a very small local minimum (with nearby local maximum) in the potential which the inflaton rolls over and generates a large spike in the matter spectrum (see also Refs.~\cite{Germani:2017bcs, Ballesteros:2017fsr}). We showed that this could be realized in a simple polynomial potential, which was compatible with all observations, including the observed spectrum on large scales.

However an important question to ask is how generic is this behavior? We showed that one of the coefficients in the polynomial, $c_5$, needs to be fine-tuned at the level of $\sim 10^{-6.5} \%$ to achieve the necessary enhancement in the power spectrum, from $\sim 10^{-9}$ to $\sim 10^{-2}$. Such possibilities may nevertheless occur occasionally in the string landscape. This may suggest that the likelihood of PBHs is small, as it is rather non-trivial to produce them in this framework. Other approaches, in which the production appears more natural, such as multi-field hybrid inflation models, have their own problems of achieving the observed spectral index on large scales, etc. It is interesting that the spike in the power spectrum scales as $\propto 1 /(\delta c_5)^2$, meaning that if $c_5$ is drawn randomly on a uniform distribution, as seems natural, then the power spectrum's probability distribution is both sharply peaked and calculable. Furthermore, since the abundance of PBHs is exponentially sensitive to the power spectrum, its probability distribution is also sharply peaked and calculable. It is relatively rare to achieve such non-trivial and calculable behavior in a landscape.

Furthermore, an important issue to address is whether there are other possible consequences of the small local minimum/maximum in the potential. While the classical field rolls over this region, quantum fluctuations can cause parts of the universe to randomly get trapped in the minimum, leading to eternal inflation. One could then produce bubble universes from tunneling out of this local minimum. While such bubbles are probably uninhabitable, since they would only have a reduced phase of subsequent inflation and so would be rather small and highly curved, it may be interesting to examine this further.

\section*{Acknowledgments}

MPH is supported by National Science Foundation grant PHY-1720332.



\end{document}